\documentstyle[12pt]{article}
\def\be{\begin{eqnarray}}
\def\ee{\end{eqnarray}}
\begin{document}

 \thispagestyle{empty}
 \begin{flushright}
 {MZ-TH/94-13}    \\
 {May 1994}           \\
 \end{flushright}
 \vspace{1cm}
 \begin{center}
 {\bf \large
New representation of two-loop propagator and vertex functions}
 \end{center}
 \vspace{1cm}
 \begin{center}
 {Andrzej Czarnecki\footnote{Address after July 1st, 1994: Institut f\"ur
Theoretische Teilchenphysik, Universit\"at
Karlsruhe, D-76128 Karlsruhe, Germany}, Ulrich Kilian}\\
  \vspace{.3cm}
 {\em Institut f\"ur Physik, Johannes Gutenberg-Universit\"at,\\
 \mbox{D-55099} Mainz, Germany}
 \end{center}
 \begin{center}
 and
 \end{center}
 \begin{center}
 {Dirk Kreimer}\\
 \vspace{.3cm}
 {\em Department of Physics, University of Tasmania, \\
      G.P.O. Box 252C, Hobart, 7001, Australia}
 \end{center}
 \hspace{3in}
\begin{abstract}
   We present a new method of calculating scalar propagator and vertex
functions in the two-loop approximation, for arbitrary masses of
particles.  It is based on a double integral representation, suitable
for numerical evaluation.  Real and imaginary parts of the diagrams are
calculated separately, so that there is no need to use complex
arithmetics in the numerical program.
 \end{abstract}

 \vspace{15mm}

 \begin{center} PACS numbers: 02.70.+d, 12.38.Bx, 11.20.Dj
 \end{center}
\newpage

\section{Introduction}
When the existence of the top quark becomes officially confirmed by
Tevatron experiments, a chapter of particle physics history will be
concluded. It has been the opinion of the vast majority of the physics
community that the top quark probably exists, as required by the
family structure of the standard model.

There is, however, no such unanimity as to what comes next.
Supersymmetry, technicolor, compositeness, are just a few  of
theoretical ideas put forward to extend the standard model, and it
will take a new generation of experiments to give us new insights.

On the other hand, when mass of the top quark becomes known, we will
have enough information to re-interpret the existing results of high precision
experiments in terms of bounds on the new physics. It will then be
possible to discard some models and concentrate on the remaining ones
in the future theoretical and experimental work. For this purpose it is
crucial to know the theoretical predictions of the standard model with
sufficient accuracy, and
in many cases this requires going to the two-loop order of
perturbation theory.

In the recent
past considerable progress has been achieved in computing massive
two-loop
Feynman diagrams
\cite{kekmc,adt93,adst93,oberamfl,fleischer94,bft93,inlo17,scharf94}.
Nevertheless, none of the existent approaches is able to give numerical
results to the three-point functions in arbitrary kinematical regions,
as needed, for example, in the determination of two-loop corrections
to partial rates of $Z$ boson decays.

The purpose of this paper is to present a solution to the problem of
calculating 2- and 3-point functions in a general mass case. In the
present work we restrict our investigations to scalar functions, which
enables us to discuss in detail the analytical properties of a given
diagram. The treatment of tensor functions will be presented in a
forthcoming paper.

Let us briefly summarize the results achieved so far. In the most
flexible approach one can represent the graph using Feynman parameters
and let a computer do the work. With sufficient memory and CPU time
available one can obtain stable numerical results.  Such an approach
is advocated by \cite{kekmc}.  The advantage of this method is that
one can treat rather general mass configurations of Feynman
diagrams. On the other hand, the accuracy which one can obtain may not
be sufficient in realistic calculations. Another problem is the
extensive use of computer power. The four- or five-fold integral
representations used by this method are on the edge of what is
available for Monte Carlo integrations nowadays.

Another approach is to use unitarity relations to find the imaginary
part of the two-loop diagram and then employ dispersion relations to
compute the real part. The first step is rather easy, since excellent
tools have been developed to calculate one-loop subdiagrams (see
e.g.~\cite{ffzp,ff}). However, the real part is
 difficult to calculate numerically, and additional complications may
arise in the presence of anomalous thresholds.

More sophisticated approaches use the idea of expanding the integrand
in some appropriate manner.  The hope is that the resulting terms are
integrable and the generated series converges.

In this line, there are two main approaches which have been described
in the literature.  Both use asymptotic expansions in some appropriate
kinematical variable, e.g.~$q^2$ or $1/q^2$.  Expanding the integrand
in this variable gives rise to simpler integrals. These might suffer
from spurious extra singularities (typically of IR type) which could
question the convergence of the generated series.  But the theory of
asymptotic expansions of Feynman diagrams, now a textbook matter
\cite{smirnov}, allows to control this problem. Results of this
approach can be found in
\cite{adt93,adst93}.

The other approach involves just a simple Taylor expansion below all
thresholds and a subsequent analytic continuation of the resulting
series. One uses a conformal mapping to enlarge the domain in which
the expansion is valid. For the 2-point case one can map the branch
cut onto a circle and the whole complex $q^2$ plane into the interior
of this circle, so that the results are valid for the whole $q^2$
domain \cite{oberamfl,fleischer94}.

One might add Pad{\'e} approximants to improve the convergence of such
series. In certain cases this method has been demonstrated to give
excellent results \cite{oberamfl,fleischer94,bft93}.

At present these two approaches have been developed only for problems
depending on one kinematical variable (e.g. 3-point function with
one mass scale and two massless external particles).
In fact, in both methods one can expect difficulties in the more general case.

The asymptotic expansions will suffer from an enormous proliferation of terms
which correct the spurious infrared behaviour, while the analytic
continuation of functions depending on several complex variables could possibly
constitute a serious mathematical problem for the second method \cite{hwa}.

Another approach consists in the expansion of some of the involved
propagators according to $x$-space or Mellin-Barnes type
representations \cite{inlo17,scharf94}. Again, one will end up with
series representations valid in certain kinematical domains.

One case which can be solved using any of the above methods is the
two-point master function of Fig.~1. While these
methods were being developed, almost all made use of yet another
representation, obtained by one of us \cite{propagd}, to check their
results. In \cite{propagd}, a very simple integral representation was
obtained, which allowed a straightforward numerical integration. One
can even use this approach to obtain results for general tensor
integrals \cite{inlo15,tensor} or to obtain integral representations
for graphs directly \cite{tensor}.

Following the ideas of \cite{propagd,vertexd} we present in this paper
an approach which is modest in its use of computer power, is of
comprehensive structure, and is valid for arbitrary kinematical
regions. The principle of our method is the same for 2- and 3- point
functions, and in the following section we explain it in much detail
with the example of the propagator function. In this way we avoid
unnecessary complications stemming from a more complex analytical
structure of the vertex.  We pay special attention to separate
calculation of real and imaginary parts.  The extension to the 3-point
functions is presented in section \ref{sec:vertex}.  In section
\ref{sec:examples} we provide two examples which illustrate
applications of our method. Both examples refer to two-loop planar
vertex functions which have been known from other approaches in some
special cases, and therefore allow a comparison of accuracy and
flexibility of various methods. The last section summarizes our
results.

\section{Propagator function}
\label{sec:propag}
An exposition of our method is simpler with the example of the
propagator function, so we are going to present it in this section in
much detail.
The two-loop propagator function is depicted in
Fig.~1 and we normalize it as in \cite{propagd,oberam}:
\be
I(q^2)=-{q^2\over \pi^4} \int d^4l \int d^4k {1\over P_1 P_2 P_3 P_4
P_5},
\label{eq:propdef}
\ee
where $q$ is the four-momentum of the external particle, $l$ and $k$ are
the internal loop momenta, and $P_i$ are the propagators of the internal
particles, as labeled in Fig.~1. We assume that the
external particle is massive (the massless limit is easy to perform),
and choose its rest frame for the calculation. The choice of parallel
and orthogonal subspaces is Lorentz invariant, and therefore the
calculation could be done in any Lorentz frame.
The four-momenta can be chosen explicitly as
\be
q^\mu&=&(q,0,0,0),\nonumber\\
l^\mu&=&(l_0+l_1,l_1,\vec l_\bot),\nonumber\\
k^\mu&=&(k_0+k_1,k_1,\vec k_\bot).
\label{eq:mom}
\ee
The splitting of the integration space $(l,k)$ into the subspaces
orthogonal and parallel to the four-momentum of the external particle was
the underlying idea of the method of doing two-loop integrals proposed
by one of us \cite{propagd,dirkphd}. Our present method differs from
that one in that we parametrize the orthogonal spaces using cylindrical
rather than spherical coordinates, which leads to very different
structure of the integrals. Such parametrization is necessary in the
case of the vertex function, as will be seen in the following section,
because there the axial coordinate  of the cylinder ($k_1$ or $l_1$)
belongs to the parallel space \cite{vertexd}. The advantage of using
cylindrical coordinates for the propagator function is that it enables
us to separate the real and imaginary parts of the function analytically
and compute them separately, so that we avoid using complex arithmetics
in the numerical program. The structure of kinematical
singularities also becomes more transparent in this parametrization.

The propagators $P_i$ of the internal particles can  be written down
using the explicit form of the four-momenta given in eq.~(\ref{eq:mom}):
\be
P_1&=& l_0^2+2l_1l_0  -l_\bot ^2 -m_1^2 +i\eta, \nonumber \\
P_2&=& (l_0-q)^2+2l_1 (l_0-q) -l_\bot ^2 -m_2^2 +i\eta, \nonumber \\
P_3&=& (l_0+k_0)^2+2(l_1+k_1)(l_0+k_0)  -l_\bot ^2-k_\bot ^2 -m_3^2
-2l_\bot k_\bot z+i\eta, \nonumber \\
P_4&=& k_0^2+2k_1k_0  -k_\bot ^2 -m_4^2 +i\eta, \nonumber \\
P_5&=& (k_0+q)^2+2k_1 (k_0+q)-k_\bot ^2 -m_5^2 +i\eta,
\ee
where $z$ is the cosine of the angle between $\vec l_\bot$ and $\vec
k_\bot$, and $-i\eta$ is a small imaginary part assigned to the masses
of internal particles.

The volume element in the integral in eq.~(\ref{eq:propdef}) can be
rewritten as
\be
d^4kd^4l= {1\over 4}dl_0 dk_0 dl_1 dk_1 ds dt d\alpha {dz \over
\sqrt{1-z^2}}, \ee
with $s\equiv l_\bot^2$ and $t\equiv k_\bot^2$, and the angle $\alpha$
describing
the absolute position of $\vec l_\bot$ and $\vec k_\bot$.
The integration over $\alpha$ is trivial and gives
an overall factor $2\pi$. Of all internal propagators only $P_3$
depends on $z$, and can be written as $P_3=A+Bz$, so that after the
integration over $z$ using
\be
\int_{-1}^1 {dz\over \sqrt{1-z^2}} {1\over A+Bz}={\pi\over \sqrt{A^2-B^2}},
\label{eq:zint}
\ee
 the propagator function becomes equal:
\be
I(q^2)&=&-{q^2\over 2\pi^2} \int {dl_0 dk_0 dl_1 dk_1 ds dt\over
\sqrt{A^2 -B^2}} {1\over P_1 P_2 P_4 P_5}
\nonumber\\
&=&-{q^2\over 2\pi^2} \int {dl_0 dk_0 dl_1 dk_1 ds dt\over \sqrt{A^2 -
B^2}} \nonumber\\
&&{1\over P_1- P_2}
{1\over P_4- P_5}
\left({1\over P_2P_5}-{1\over P_2P_4}-{1\over P_1P_5}+{1\over P_1P_4} \right)
\nonumber\\
\ee
In the next step we integrate over $k_1$ and $l_1$ using the Cauchy
theorem. The Fubini theorem allows the change of the order of integration, as
the integral over the modulus of the integrand exists. This is true
even in the degenerate limit $q\rightarrow 0$ where $(P_1 - P_2)$ and
$(P_4 - P_5)$ are independent of any integration variables.
In order to determine which terms contribute we consider the
analytical structure of the integrand. As explained in
ref.~\cite{vertexd}, the position of cuts of the square root
$\sqrt{A^2-B^2}$
as a function of $l_1$ and $k_1$ is determined by the sign of $l_0+k_0$;
they are either both in the upper half-planes or both in the lower ones,
for $l_0+k_0<0$ and $l_0+k_0>0$ respectively. Let's  first  take
$l_0+k_0>0$, so that we have to close the contours of $k_1$ and $l_1$
integrations in the upper half-planes. The following propagators have
poles in this half-plane: $P_1$ if $l_0<0$, $P_2$ if $l_0-q<0$,
$P_4$ if $k_0<0$, and
$P_5$ if $k_0+q<0$. Not all of these conditions can be reconciled with
the inequality $l_0+k_0>0$, and in fact only the term $1/(P_2P_4)$
contributes. Analogously, for $l_0+k_0<0$, only  $1/(P_1P_5)$
contributes.
Moreover, in each case the  three inequalities which must be satisfied
restrict the region of integration to
a triangle in the $(k_0,l_0)$ plane. Integration over the rest of the
plane gives identically zero. The relevant triangles for  positive
and negative $l_0 +k_0$  are depicted in Fig.~2(a)
and (b), respectively.

After the $k_1$ and $l_1$ integrations the propagator function is
represented by:
\be
I(q^2)&=&-{q^2\over  2}
\left(\int\!\!\!\int_{T_a}{1\over k_0(l_0-q)}
+ \int\!\!\!\int_{T_b}    {1\over l_0(k_0+q)}\right) dl_0 dk_0
 \int_0^\infty\! ds \int_0^\infty\! dt
\nonumber\\
&&{1\over P_1- P_2}
{1\over P_4- P_5} {1\over \sqrt{A^2-B^2}} \nonumber \\
&=&{1\over 2}
\left(\int\!\!\!\int_{T_a} + \int\!\!\!\int_{T_b}\right) dl_0 dk_0
 \int_0^\infty\! {ds\over s+s_0-i\eta} \int_0^\infty\! {dt\over t+t_0-i\eta}
\nonumber\\
&& {1\over \sqrt{(at+b+cs)^2-4st}},
\label{eq:gen}
\ee
with $T_a$ and $T_b$ representing the triangles depicted in
Fig.~2, and $a$, $b$ and $c$ being functions of masses
$m_i$ and momenta
$k_0$ and $l_0$. These functions, although in general quite involved, can be
easily calculated with any algebraic manipulation program by
inserting expressions for the position of poles in $l_1$ and $k_1$ into
the quantity $A$ obtained from $P_3$ ($=A+Bz$). Of course, these
coefficients, as well as explicit forms of the propagators $P_i$ have to be
calculated separately for each of the triangles $T_a$ and $T_b$. We
give the explicit formulas in the Appendix; the only important fact we
need to now for further derivations is that the coefficient
functions $a$ and $c$ are always negative in the domain of integration.
Finally, the structure of quantities $s_0$ and $t_0$ is essential for the
calculation of the imaginary part of the self-energy diagram. Their
explicit form will be given in the following section.

\subsection{Imaginary part of the propagator function}
In equation~(\ref{eq:gen}) one integrates only over positive values of
$s$ and $t$, so the contributions to the imaginary part
 occur either if $s_0$ or $t_0$ are negative, or if the argument of the
square root becomes negative. These two possibilities correspond to
 cuts of the diagram in Fig.~1 across two and three
internal lines, respectively. We first discuss the cut across two
internal lines. In the general mass case, the explicit forms of $s_0$
and $t_0$ are:
\be
s_0&=&l_0^2+{m_2^2-m_1^2-q^2\over q}l_0+m_1^2,
\nonumber\\
t_0&=&k_0^2-{m_5^2-m_4^2-q^2\over q}k_0+m_4^2.
\ee
The condition that $s_0$ ($t_0$) be negative is a quadratic inequality
for $l_0$ ($k_0$). The solutions describe stripes in the $(k_0,l_0)$
plane in which the integral gets an imaginary part. These regions are
depicted in Fig.~3. It can be seen that
real solutions for $l_0$ and $k_0$ exist if $q$ satisfies
\be
q^2 &>& (m_i+m_j)^2 ,
\label{eq:thresh}\\
\mbox{\rm or}\qquad q^2 &<& (m_i-m_j)^2 ,
\label{eq:psthresh}
\ee
where $i=1$, $j=2$ for $s_0$, and $i=4$, $j=5$ for $t_0$.
The  condition (\ref{eq:thresh}) describes a normal threshold
corresponding to a cut across the lines 1 and 2 or 4 and 5 in the
diagram in Fig.~1, whereas (\ref{eq:psthresh})
corresponds to a pseudothreshold; it can be seen that it leads to
solutions for $k_0$ and $l_0$ outside the integration triangles.

If $s_0$ or $t_0$ are negative we rewrite the integral representation
in the form:
\be
\mbox{Im}\, I(q^2)&=&{\pi\over 2}
\left(\int\!\!\!\int_{T_a} + \int\!\!\!\int_{T_b}\right) dl_0 dk_0
%\nonumber\\&&
\left\{
{\rm P.V.}  \int_0^\infty\! {ds\over s+s_0}  \int_0^\infty\! dt \delta(t+t_0)
\right.
\nonumber\\&&
\left.
+{\rm P.V.}  \int_0^\infty\! {dt\over t+t_0}  \int_0^\infty\! ds \delta(s+s_0)
\right\}
{1\over \sqrt{(at+b+cs)^2-4st}}.
\ee
In each case  one integration cancels a delta function and the
other is elementary and can be done with Euler's change of variables
\cite{fichte}.

In order to analyze the contribution to the imaginary part from the square
root in eq.~(\ref{eq:gen}), we have to find the four-dimensional
region in variables $(s,t,l_0,k_0)$ where the function
$(at+b+cs)^2-4st$ is negative. The projection of
such region on the $(s,t)$ plane is described by an ellipse:
\be
s={1\over c^2}\left[\sqrt{t}\pm \sqrt{t(1-ac)-bc}\right]^2,
\ee
from which it follows that the condition for the imaginary part to
occur in this case  is
$b>0$ (the expression $1-ac$ is always negative in the region of
integration).  By an explicit calculation we convince ourselves that
the inequality $b>0$ describes a closed region inside the triangular
regions of integration (see Fig.~3), provided that
at least one of
the threshold conditions $q^2>(m_1+m_3+m_5)^2$ or  $q^2>(m_2+m_3+m_4)^2$
are satisfied. As already indicated, they correspond to the three
particle intermediate states,  or cuts across the lines 1, 3, 5 or 2,
3, 4. Also in this case there exist pseudothreshold solutions
beyond the limits of the integration region.

The integrations over $s$ and $t$ in this case can be done by
observing that the square root in the integrand vanishes on the
boundary of the integration region, and one can therefore use the
formula analogous to eq.~(\ref{eq:zint}).

Finally, we note that if $m_3$=0, both types of imaginary
contributions are divergent. Their sum, however, is free from this
infrared divergence.

\subsection{The real part}
The calculation of the real part is conceptually simpler, since the
general formula is just:
\be
\lefteqn{\mbox{Re}\, I(q^2)=}
\nonumber\\
& &{1\over 2}
\left(\int\!\!\!\int_{T_a} + \int\!\!\!\int_{T_b}\right) dl_0 dk_0
%\nonumber\\&&
\mbox{P.V.}  \int_0^\infty\! {ds\over s+s_0}
\mbox{P.V.}  \int_0^\infty\! {dt\over t+t_0}
{1\over \sqrt{(at+b+cs)^2-4st}}
\nonumber\\& &
-{\pi^2\over 2}
   \int\!\!\!\int_V dl_0 dk_0 {1\over \sqrt{(at_0-b+cs_0)^2-4s_0t_0}},
\ee
where the second term  originates from the
region $V$ of the $(l_0,k_0)$ plane where
we get a product of two imaginary quantities, yielding a contribution
to the real part. This term is only present if $q^2$ is above at least
two thresholds, and the  boundary of $V$ can be determined from the
condition that two imaginary contributions occur there. An example of
such region can be seen in Fig.~3 as a triangle formed by a
cross-section of two stripes corresponding to the two-particle cuts.

In the first term we have to perform a double integration over $s$
and $t$.  Two consequtive Euler's changes of
variables reduce the problem to  simple integrals of the form
\be
\int {\ln x \over x^2 +\alpha x +\beta} dx,
\ee
which can be expressed in terms of dilogarithms \cite{lewin}.

\section{Vertex function}
\label{sec:vertex}
In the standard model calculations one encounters  two topologies of
two-loop vertex functions, as depicted in Fig.~4. In
the present paper we  consider only the planar case. The crossed
diagram can be calculated using the same technique \cite{vertexd};
details and numerical results will be published separately.

Our approach to the calculation of the vertex function follows closely the
method presented in the previous section in the context of the propagator. We
present some details with the example of a decay of a particle of mass
$\sqrt{q^2}$ into particles with four-momenta $q_1$ and $q_2$.
For fixed values of $q_1^2$ and $q_2^2$ we define the vertex function as:
\be
V(q^2)=\int{{\rm d}^4k{\rm d}^4l\over P_1P_2P_3P_4P_5P_6}.
\label{eq:vertnorm}
\ee
With external momenta parametrized as
\be
q_1&=&(q_{10},q_z,0,0), \nonumber \\
q_2&=&(q_{20},-q_z,0,0),
\ee
and the internal ones -- as in the eq.~(\ref{eq:mom}), the propagators
$P_i$ have the following form:
\be
P_1&=&l_0^2+q_{10}^2-q_z^2+2 l_1 (l_0+q_{10}-q_z)+2 l_0 q_{10}-l_\bot^2-m_1^2
+i\eta,\nonumber\\
P_2&=&l_0^2+q_{20}^2-q_z^2+2 l_1 (l_0-q_{20}-q_z)-2 l_0 q_{20}-l_\bot^2-m_2^2
+i\eta,\nonumber\\
P_3&=& (l_0+k_0)^2+2(l_1+k_1)(l_0+k_0)  -l_\bot ^2-k_\bot ^2 -m_3^2
-2l_\bot k_\bot z+i\eta, \nonumber \\
P_4&=&k_0^2+q_{10}^2-q_z^2+2 k_1 (k_0-q_{10}+q_z)-2 k_0 q_{10}-k_\bot^2-m_4^2
+i\eta,\nonumber\\
P_5&=&k_0^2+q_{20}^2-q_z^2+2 k_1 (k_0+q_{20}+q_z)+2 k_0 q_{20}-k_\bot^2-m_5^2
+i\eta,\nonumber\\
P_6&=&k_0^2+2 k_1 k_0-k_\bot^2-m_6^2+i\eta.
\ee
Exactly as in the calculation of the propagator function, the angular
integrations are done first, yielding the same type of a square root
function, whose cuts determine how to close contours of $l_1$ and
$k_1$ integrations. The only difference is that there are 6
propagators now and the partial fraction becomes slightly more
complicated. Consequently, there are more terms which give nonzero
contributions to the integral, and  one has to distinguish four
different integration domains in the $(k_0,l_0)$ plane. They are
depicted in Fig.~5, and we will label them
accordingly $T_{a,b,c,d}$.
The representation of the vertex function before splitting it into the
imaginary and real parts can be written down in analogy to eq.~\ref{eq:gen}:
\be
V(q^2)&=&
\left(\int\!\!\!\int_{T_a} + \int\!\!\!\int_{T_b}
+ \int\!\!\!\int_{T_c}+ \int\!\!\!\int_{T_d}\right) dl_0 dk_0
C(k_0,l_0)
\nonumber\\ &&
 \int_0^\infty\! {ds\over s+s_0-i\eta} \int_0^\infty\! {dt\over
(t+t_0-i\eta)(t+t_0^\prime-i\eta)
}
{1\over \sqrt{(at+b+cs)^2-4st}},\nonumber\\
\label{eq:genver}
\ee
where $C(k_0,l_0)$, $s_0$, $t_0$,  $t_0^\prime$, $a$, $b$, $c$ are
rational functions of $l_0$ and $k_0$, which have to be computed in
each of the four integration regions $T_i$ separately.
The only difference from the analogous representation of the two-point
function consists in an extra $t$-dependent term in the denominator,
which arises because there are now four propagators in one of the
loops, instead of three in the previous case. This new term does not
cause any difficulties since we can decompose the integrand into
simple fractions. The calculation of real and imaginary parts can now
be done in the same manner as described in the previous section.

The appearance of an extra $t$-dependent term corresponds to the new
way of cutting the diagram, across the lines 4 and 6 or 5 and 6, and
reflects the richer analytical structure of the vertex diagram. Also
the fact that there are 4 domains of integration in the present case
is connected with four possible 3-particle cuts of the planar vertex.
The square root in the integrands reflects in every domain presence of
a distinct 3-particle threshold. This result is quite general (e.g., there
are 2 domains in case of a 2-point function and 6 in the 3-point
crossed vertex).

Finally we note that $s$ and $t$ integrations can be done analytically
just like in the case of the 2-point function, so that we obtain a
double integral representation for the vertex function.

\section{Examples}
\label{sec:examples}
The method which we have presented in this paper can be applied to the
calculation of any two-loop two- or three- point function. In the
present section we give examples of two classes of planar vertex
functions for which results have been known in some limiting cases,
and therefore it is possible to verify the computation. Both examples
will also show that our method significantly extends the class of
computable diagrams or improves the accuracy of the result.

As the first example we take a diagram of Fig.~4(a)
with $q_1^2=q_2^2=0$, $m_1=m_2=m_4=m_5=m_6\equiv M$, and study
the real and imaginary parts of the corresponding amplitude at various
values of $q^2$ and $m_3$.
For the particular value of $m_3=0$ this
diagram has been studied in refs.~\cite{kekmc,oberamfl,fleischer94},
so this limiting case is well known.
The physical motivation for
calculating this diagram is, e.g., the decay of a Higgs particle
into
two photons via a heavy quark triangle.
The massless particle on line
3 represents a gluon. With the help of our method we can
calculate this diagram for nonvanishing $m_3$, corresponding, for
example, to an exchange of another Higgs particle inside the quark
loop. The main complication with respect to the massless case is that
there are now two thresholds on the positive real axis of variable $q^2$,
corresponding to cuts across two and three internal lines. In fact
the case of vanishing $m_3$ is the hardest one to compute numerically,
because the contributions from both cuts are infrared divergent. The
limit is, however, approached smoothly, and one obtains excellent
agreement with previous calculations already for $m_3/M=10^{-3}$,
for which results are shown in Fig.~6 (the real part) and
in Fig.~7 (the imaginary part)
as a solid line.
The circles denote the points for which the results have been
published in \cite{fleischer94}, and the long- and short- dashed lines
depict the cases of $m_3$ equal to $M$ and $2M$, respectively.
For easy comparison with
refs.~\cite{kekmc,oberamfl,fleischer94}\footnote{Ref.
\cite{oberamfl,fleischer94} omitts the factor $10^9$}
we take $M=150$ GeV, and plot the function $I(q^2)$ defined as:
\be
I(q^2)=10^9 \pi^4 V(q^2),
\ee
with $V(q^2)$ defined in eq.~\ref{eq:vertnorm}.
As a cross check we have used a program based on dispersion relations and
got excellent agreement, although the calculation of the real part is
much more time consuming in the dispersion approach.

In our second example we examine a scalar diagram corresponding to a
decay of a particle of mass $\sqrt{q^2}$ into two heavy quarks, with
two massive $Z$ bosons exchanged between the outgoing quarks.
In the notation of Fig.~4(a) we take
$\sqrt{q_1^2}=\sqrt{q_2^2}=m_1=m_2=m_4=m_5\equiv M$ and
$m_3=m_6=m_Z$, where $m_Z$ is the mass of the $Z$ boson which we
take equal 91.17 GeV. In this process the outgoing particles are
massive, and no results have been obtained in such cases in the
approaches using asymptotic expansions.  Therefore we compare our
results with the numbers obtained using Monte Carlo approach.
 The real part of this diagram is depicted in
Fig.~8 for two masses of the heavy quark: $M=$ 150 GeV
(solid line) and 174 GeV (dashed line). The dots represent data points
of ref.~\cite{kekmc}\footnote{We are very grateful to Dr. Fujimoto
for sending us his numerical results.} obtained for $M=$ 150 GeV.
Similarly, the imaginary part is plotted in Fig.~9.
Our results
lie within error bars indicated by the authors of ref.~\cite{kekmc},
our numerical error is, however, at least two orders of magnitude smaller.
We have compared our results
with numbers given by unitarity relations and obtained agreement up to
one part in $10^6$, and even further refinement is possible.

\section{Conclusions}
In this paper we have presented a method of calculating two-loop two-
and three-point functions, based on a double integral representation
convenient for numerical evaluation.  Although this method is
applicable to arbitrary mass configurations, we introduced it
by presenting relatively simple examples.  Already these examples show
that our method allows treating more general cases and obtain better
accuracy than has been possible using other methods.

There are three directions in which our results can be generalized.
First, one should consider mass cases in which other types of
thresholds are present, and also include the crossed topology for the
three-point functions (Fig.~4).  This demands a more
detailed investigation of the domains contributing to the imaginary
part, but does not pose a principal problem.

Second, one can generalize this method easily to the case of
four-point functions. Just as in the transition from two-point to
vertex functions, the transition from vertex functions to box
functions results in a proliferation of terms, while the general idea
remains unchanged.  Certainly, the three-dimensional parallel space
requires an appropriate treatment, but the general structure
of the integrands will not be altered.  We will present these results
elsewhere.

However, even when all scalar integrals are available, this is not the
end of a practical two-loop calculation, and this brings us to the
third line of generalizations. One still has to treat the tensor
structure of a given graph; its subdivergent behaviour has to be taken
into account and one-loop counterterm diagrams have to calculated; and
one has to impose renormalization conditions which typically involve
on-shell singularities.

In recent works \cite{tensor,adelaide} one of us proposed a way of
achieving all these aims using a concise algebraic algorithm. This
algorithm is now being tested with simple examples, and we intend to
demonstrate its use in standard model calculations in the
future. Together with the results on UV-finite scalar two-loop
functions obtained here, this paves the way to a convenient approach
to two-loop Feynman graph calculations, especially for the standard
model, where one is usually plagued by enormous tensor-algebraic and
analytical difficulties, resulting from the interaction of fields with
various masses and Lorentz properties.

\section*{Acknowledgment} We wish to express thanks to  D. Broadhurst,
J.G. K\"orner and K. Schilcher for interest in this work,
encouragement, and many helpful discussions.  We gratefully
acknowledge helpful comments from K.G. Chetyrkin, A.I. Davydychev,
P. Findeisen, M. Je\.zabek, A.N. Kamal, G.J. van Oldenborgh,
D. Pirjol and J.B. Tausk.
We also wish to thank J. Fleischer and O.V. Tarasov for communicating
pre-publication results of ref.~\cite{oberamfl,fleischer94}, and J.
Fujimoto for sending us numerical results used in ref.~\cite{kekmc}.
We thank G.J. van Oldenborgh for kind permission to use parts of his
program library FF for evaluation of dilogarithms and Clausen functions.

A.C. was supported by a grant from DFG and by Graduiertenkolleg of the
Univ. of Mainz and U.K. by Cusanuswerk.

\section*{Appendix}
We list here explicit formulae for the coefficient functions $a$, $b$
and $c$, as defined in eq.~(\ref{eq:gen}), for the case of the
2-point function. There are two  cases which we have to treat
separately, depending on the residues of the propagators from which
$k_1$ and $l_1$ were determined:

\noindent
\underline{Case $P_1P_5$}
\be
a &   =&{l_0 - q \over k_0 + q},\nonumber\\
b   & =& {1\over l_0(k_0+q) }
\left[(m_1^2-q l_0)(k_0^2+k_0 l_0 +k_0 q+l_0 q)\right.\nonumber\\
&&\left.\qquad\qquad
-m_3^2l_0(k_0+q)+m_5^2l_0(k_0+l_0) \right]+i\eta,\nonumber\\
c   &=&{k_0\over l_0}.
\ee

\noindent
\underline{Case $P_2P_4$}
\be
a     &=&{l_0\over k_0},\nonumber\\
b&=&{1\over k_0(l_0-q)}
\left[(m_4^2+k_0 q)(l_0^2+k_0 l_0-k_0 q-l_0 q)\right.\nonumber\\
&&\left.\qquad\qquad+m_2^2k_0(k_0+l_0)
+m_3^2 k_0 (q-l_0)\right]+i\eta,\nonumber\\
c &   =&{k_0 + q \over l_0 - q}.
      \ee

\section*{Figure captions}
\begin{itemize}
\item[]{Fig.~1: Two-loop propagator ``master'' diagram}
\item[]{Fig.~2: Regions of integration in the $(k_0,l_0)$ plane}
\item[]{Fig.~3: Example of the structure
of the integration region in the $(k_0,l_0)$
plane: term arising from the residues in $P_1$ and $P_5$,
for $m_1=m_2=m_4=m_5$. The shaded areas correspond to
contributions of 2-particle cuts to the imaginary part. The irregular
closed region depicts the area where the 3-particle state contributes. }
\item[]{Fig.~4: Two topologies of two-loop vertex diagrams}
\item[]{Fig.~5: Four integration domains for the vertex function}
\item[]{Fig.~6: Real part of the two-loop planar vertex function for $m_i=M$
($i=1,2,4,5,6$),
 $m_3$
vanishing (solid line), $m_3=M$ (long dash), and $m_3=2M$ (short dash).}
\item[]{Fig.~7: Imaginary part of the
two-loop planar vertex function for $m_i=M$
($i=1,2,4,5,6$), $m_3$
vanishing (solid line), $m_3=M$ (long dash), and $m_3=2M$ (short dash).}
\item[]{Fig.~8: Real part of the two-loop planar vertex function for
$\protect\sqrt{q_1^2}=\protect\sqrt{q_2^2} =m_1=m_2=m_4=m_5=M$
and $m_3=m_6=m_Z$, for $M=150$ GeV (solid line), and $M=174$ GeV
(dashed line).}
\item[]{Fig.~9: Imaginary part of the two-loop planar vertex function for
$ \protect\sqrt{q_1^2}=\protect\sqrt{q_2^2}=
m_1=m_2=m_4=m_5=M$
and $m_3=m_6=m_Z$, for $M=150$ GeV (solid line), and $M=174$ GeV
(dashed line).}
\end{itemize}

\end{document}